# Electrical evolution of W and WC Schottky contacts on 4H-SiC at different annealing temperatures

Marilena Vivona[1], Gabriele Bellocchi[2], Raffaella Lo Nigro[1], Simone Rascunà[2], Fabrizio Roccaforte[1]

[1] CNR-IMM, Strada VIII n.5 Zona Industriale, I-95121, Catania, Italy
[2] STMicroelectronics, Stradale Primosole 50, I-95121, Catania, Italy

E-mail: marilena.vivona@imm.cnr.it

**Abstract**

In this paper, we investigate the electrical evolution of tungsten (W) and tungsten carbide (WC) Schottky contacts on 4H-SiC subjected to thermal treatments at different annealing temperatures from 475 to 700 °C. For each annealing temperature, the uniformity of the Schottky barrier height ($\Phi_B$) and ideality factor (n) was monitored by current-voltage (I–V) measurements in forward bias, performed over sets of equivalent diodes. Good values of n (below 1.05) were found for both contacts up to thermal annealing at 700°C. On the other hand, the barrier of the two contacts behaves differently. For the W/4H-SiC diode, the $\Phi_B$ increases with the annealing temperature (from 1.14 eV at 475°C to 1.25 eV at 700°C), whereas the Schottky barrier in WC/4H-SiC features a slight reduction already with thermal annealing at 475 °C, remaining almost constant at around 1.06 eV up to annealing at 700°C. A deeper characterization was performed on the 700°C-annealed contacts by studying the temperature-dependence of the Schottky parameters by current-voltage-temperature (I-V-T) characterization. The $\Phi_B$ and n behaviour with temperature indicates the presence of a nanoscale lateral inhomogeneity for both Schottky contacts, which can be described by Tung's model. Finally, the temperature-dependence of the reverse characteristics could be described by the thermionic field emission model (TFE), accounting for the temperature dependent barrier height determined from forward characterization.

## 1. Introduction

For the last two decades, Schottky barrier diodes (SBDs) on hexagonal silicon carbide (4H-SiC) have been present on the market, finding increasingly number of applications in several fields, such as power supplies in consumer electronics, energy conversion systems in renewable sources and hybrid/electric vehicles, industrial motors control and so on [1,2,3]. In fact, due to the excellent properties of 4H-SiC, such as a wide bandgap and a high critical electric field [3], these devices can operate at higher voltage and current levels and with lower on-resistance with respect to the traditional silicon diodes [4]. Consequently, they can guarantee a lower power consumption and better energy efficiency. Currently, a large variety of 4H-SiC SBDs operating in the range of 600-3000V at current levels up to 100A are available on the market [2]. Nevertheless, margins of improvement of the device performances are still possible, achievable through the control and optimization of the metal/4H-SiC Schottky barrier and a full understanding of the current transport mechanisms at this interface [4].

The core of a Schottky diode is the metal/semiconductor interface, whose electrical properties rule the current transport through the contact and thus the overall device behaviour [5]. For that reason, metal/4H-SiC interfaces have been extensively investigated over the last three decades, with particular attention to the choice of the metal and its evolution in the Schottky contact formation [5,6,7,8,9].

In this context, titanium (Ti) and nickel silicide ($Ni_2Si$) exhibit highly reproducible Schottky barrier height values on 4H-SiC of 1.25 and 1.65 eV, respectively [10]. Hence, these metals have been typically addressed to different applications. In fact, the lower barrier of Ti enables to fulfil the requirement of a low forward voltage drop $V_F$ in power electronics applications [5]. On the other hand, $Ni_2Si$, with its higher barrier, gives a lower leakage current as often required in sensors and detectors technologies [11,12].

Recently, special attention has been paid to those metallization schemes containing metal with low work-function and offering good thermal stability, such as molybdenum (Mo) and tungsten (W) [13,14,15,16,17] and their compounds [18,19,20]. As a matter of fact, metals with low work function can guarantee a minimization of the on-state conduction losses, making this configuration highly aimed in industrial 4H-SiC-based Schottky device development [1,21,22]. However, a certain degree of inhomogeneity has been observed for this kind of contact. The inhomogeneity character has been often interpreted by the Tung's model [23], which assumes a nanoscale lateral inhomogeneity of the Schottky barrier described by the presence of regions with low-barrier embedded in a ideal uniform high-barrier background. In particular, *Renz et al.* [21] recently studied the improvement of Mo/4H-SiC Schottky barrier upon appropriate surface annealing conditions.

In this context, it is also particularly interesting to study the effect of the annealing temperature on the evolution of the Schottky barrier height on low work-function metals on 4H-SiC Schottky diodes.

This paper systematically investigates the evolution of the electrical properties of tungsten (W) and tungsten carbide (WC) Schottky contacts to 4H-SiC subjected to thermal annealing in the range 475-700°C. Specifically, the temperature-dependence of the Schottky barrier height $\Phi_B$ and the current transport mechanisms are studied under both forward and reverse bias in selected annealed samples to gain insights into the inhomogeneous nature of the Schottky barrier.

## 2. Experimental details

Commercial n-type 4H-SiC wafers, with a 9.5μm-thick epitaxial layer and nominal doping concentration of $N_D = 8 \times 10^{15}$ cm$^{-3}$, were used for our study. Before the front-side processing, a large-area Ohmic contact was fabricated on the back-side of the wafer by Ni deposition followed by rapid thermal annealing (RTA) at 950°C in $N_2$ [24]. On the front-side, Schottky diodes with an active contact area $A = 4.53 \times 10^{-2}$ cm$^2$ were defined by means of optical lithography and lift-off processing steps. Barrier material, namely 80nm-thick tungsten (W) or tungsten carbide (WC) layer, was deposited by DC magnetron sputtering. The Schottky diodes were then subjected to different thermal annealing treatments for 10 min in $N_2$-atmosphere, at temperatures ranging from 475 to 700°C. The as-deposited and thermal annealed Schottky contacts were electrically characterized by means of current-voltage (I–V) measurements, performed in a Karl-Suss MicroTec probe station equipped with a parameter analyser. The electrical parameters of the contacts (ideality factor *n* and Schottky barrier height $\Phi_B$) were averaged over a set of measurements on several equivalent diodes. For selected W/4H-SiC and WC/4H-SiC annealed contacts, X-ray diffraction (XRD) analysis was carried out in grazing incident mode (0.5°) using a Smartlab Rigaku diffractometer equipped with a rotating anode of Cu K$_\alpha$ radiation operating at 45kV and 200 mA. Furthermore, the temperature-dependence of the electrical characteristics (I–V–T) of these diodes was studied under both forward and reverse bias with measurement temperature varying in the range 25–125 °C.

## 3. Results and Discussion

Firstly, the evolution of the forward current-voltage characteristics of the W/4H-SiC and WC/4H-SiC Schottky diodes was monitored for the as-deposited devices and for the devices annealed at different temperatures from 475°C up to 700°C. The representative current density-voltage (J–V) curves for each case are reported in a semilog plot in Figs. 1a and 1b for the W and WC/4H-SiC Schottky diodes, respectively. As can be seen, the two contacts behave in a different manner. In particular, a gradual shift toward higher voltage (increase of the turn-on voltage) of the J–V characteristics with the increase of the annealing temperature is observed in the W/4H-SiC contact (Fig. 1a). On the other hand, the WC/4H-SiC contacts showed a negative shift with respect to the as-deposited device up to the annealing temperature of 700 °C (Fig. 1b). A further increase of the annealing temperature at 800°C (J-V curves not shown), produced in both contacts a significant increase of the series resistance, thus suggesting that a degradation of the metal/4H-SiC interface likely occurs.

In order to extrapolate the relevant electrical parameters of the contact, we analysed the forward characteristics by considering the thermionic emission (TE) model [25]. Accordingly, the ideality factor *n* and Schottky barrier height $\Phi_B$ can be derived by fitting the linear region of the semilog plot of the forward curves with the relationship [26]:

$$I = AA^*T^2 exp\left[\left(-\frac{q\phi_B}{k_BT}\right)\right] exp\left[\left(\frac{qV_F}{nk_BT}\right)\right] \qquad (1)$$

where A is the diode area, A* is the effective Richardson constant of 4H-SiC (146 A×cm$^{-2}$×K$^{-2}$) [27], T is the absolute temperature, q is the elementary charge, k$_B$ is the Boltzmann constant and V$_F$ is the applied forward voltage.

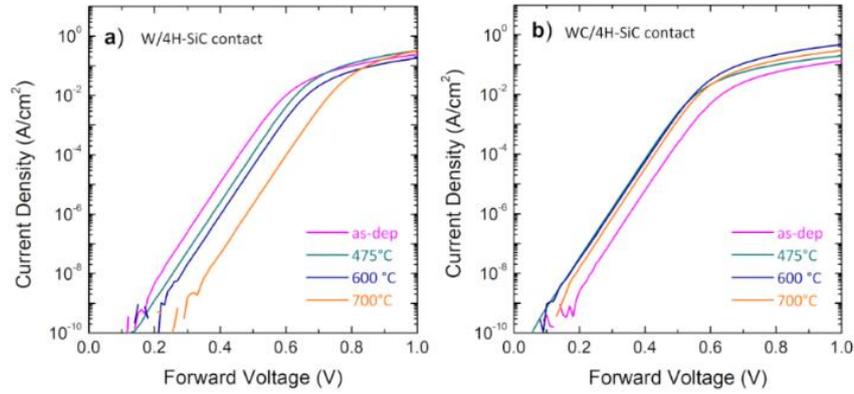

**Figure 1**. Forward current density-voltage (J-V) charcateristics for *a)* W/4H-SiC and *b)* WC/4H-SiC Schottky diodes for as-deposited devices and subjected to thermal annealing at temperatures ranging from 475 to 700 °C

This I-V analysis has been performed for each annealing temperature over sets of about 30 equivalent diodes distributed over the wafer surface, in order to determine the averaged electrical properties of the Schottky contacts in the as-deposited and under different annealing conditions.

Fig.2 reports the average value of the ideality factor *n* for both the W/4H-SiC and WC/4H-SiC contacts as a function of the annealing temperature. The associated error bars are given by the standard deviation. As can be seen, for both contacts the ideality factor remains almost constant between 1.02 and 1.03 up to annealing at 700°C.

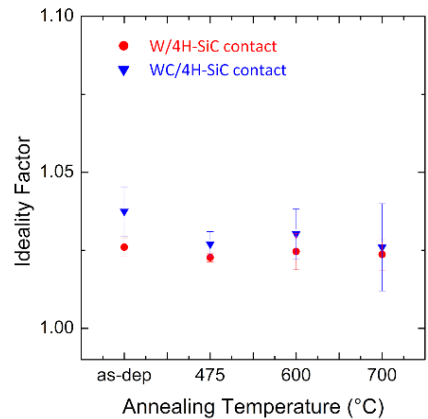

**Figure 2**. Average ideality factor *n* derived from sets of forward I-V measurements on equivalent W/4H-SiC and WC/4H-SiC Schottky diode sets for as-deposited and annealed contacts at 475, 600, and 700°C.

For both contacts, narrow statistical distributions (within ~1%) were observed for the barrier height for all annealing temperatures, indicating a good reproducibility of the W and WC barrier formation on 4H-SiC. As an example, the statistical distributions of the Schottky barrier height Φ$_B$ values for as-deposited and samples annealed at 700°C are illustrated in Figs. 3a and 3b for the W/4H-SiC and Figs. 3c and 3d for the WC/4H-SiC contact. Interestingly, more symmetric (Gaussian-like) distributions of the measured barrier heights values were observed in the annealed WC/4H-SiC with respect to the W/4H-SiC diode.

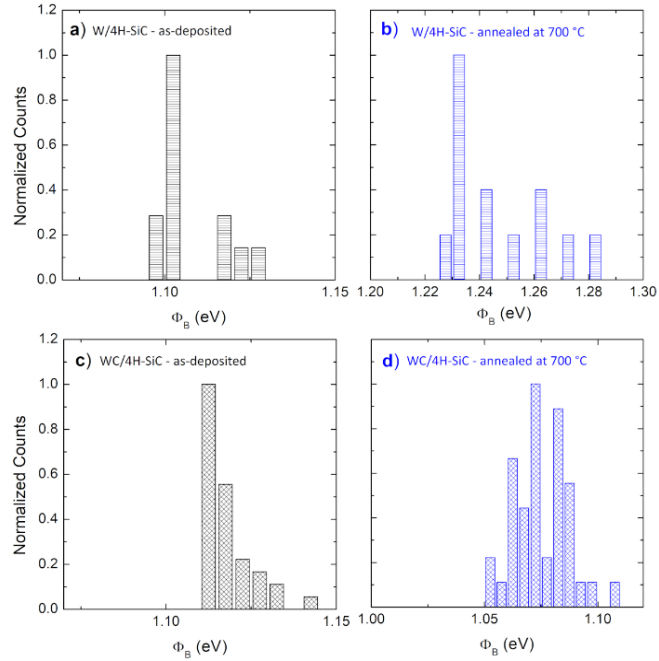

**Figure 3**. Statistical distributions of the barrier height values derived from I-V measurements on *(a-b)* W/4H-SiC and *(c-d)* WC/4H-SiC Schottky diodes after deposition and annealing at 700°C.

The appearance of asymmetric barrier height distributions observed in the as-deposited contacts and at low annealing temperatures can be ascribed to the occurrence of a Fermi-level pinning, due to the presence of surface defects [28,29]. Indeed, in the two as-deposited contacts, similar values of the barrier height are measured, thus suggesting that the ideal Schottky-Mott relation is not satisfied [25], but rather a pinning of the Fermi level is occurring [30]. On the other hand, as we have already reported in Ref. [20], the barrier height distribution of the WC/4H-SiC contact annealed at 700°C can be fitted by a Gaussian curve centred at 1.075 eV and with a standard deviation of 0.013 eV.

The mean value for the barrier height and standard deviation obtained from this statistical analysis are summarized in the plot of Fig.4. Evidently, with the increase of the annealing temperature, a gradual increase of the barrier up to 1.25 eV is observed for the W/4H-SiC diode. On the other hand, for the WC/4H-SiC diode, the barrier exhibited a slight reduction after annealing processes up to 700 °C. It is worth mentioning that the WC/4H-SiC contact shows a low value of ideality factor (1.03) and barrier height (1.06 eV) already for an annealing temperature of 475°C.

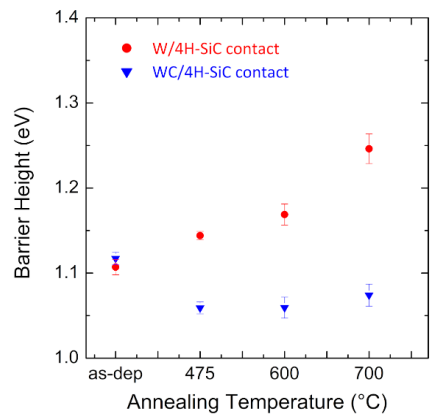

**Figure 4**. Average Schottky barrier height $\Phi_B$ for the as-deposited and annealed contacts at 475, 600 and 700 °C, extracted from the statistical analysis on I-V dataset.

In order to gain insights on the Schottky barrier nature and current transport mechanism through the interfaces, we monitored the temperature-dependence of the I-V characteristics for the W/4H-SiC and WC/4H-SiC Schottky diodes. This analysis has been carried out on the samples annealed at 700 °C, which was the maximum annealing temperature before electrical degradation of the contacts. At this annealing temperature, the XRD analysis, shown in Fig.5, indicated the presence of metallic W cubic phase for the W/4H-SiC and hexagonal $W_2C$ phase for the WC/4H-SiC contacts. Both samples possess polycrystalline structure.

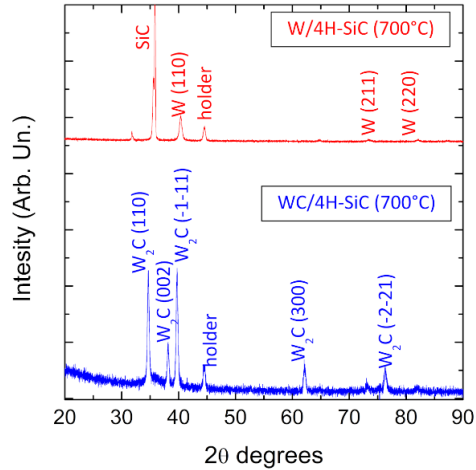

**Figure 5**. XRD patterns of the 700°C-annealed W/4H-SiC and WC/4H-SiC contacts.

For both contacts, the related forward J-V characteristics, acquired at measurement temperature varying in the range 25-125 °C with step of 25 °C, are depicted in Fig. 6.

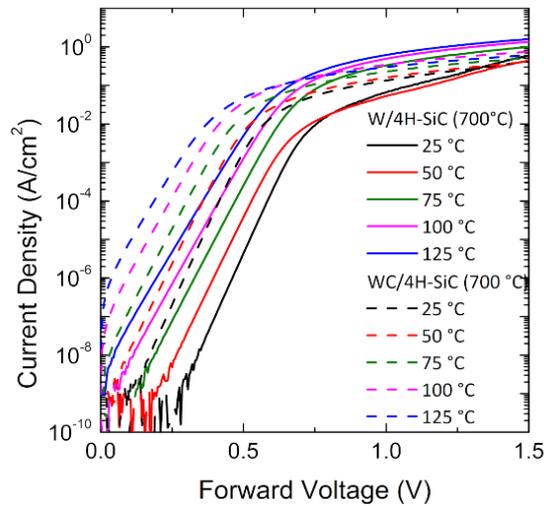

**Figure 6**. Forward current density-voltage charcateristics measured at different temperatures in the range 25-125°C for W/4H-SiC and WC/4H-SiC contacts annealed at 700 °C.

According to the TE model, for both diodes the forward current increases with increasing the measurement temperature. Figs. 7a and 7b report the ideality factor $n$ and barrier height $\Phi_B$ values as a function of the measurement temperature. For both W/4H-SiC and WC/4H-SiC annealed contacts, we observed a decrease of the ideality factor n with increasing the measurement temperature (Fig. 7a), and an increase of the barrier height $\Phi_B$ (Fig. 7b).

In spite of the low values (close to one) of the ideality factor, the observation of such a temperature-dependence of n and $\Phi_B$ is an indication of a deviation from a pure TE model. Hence, a certain degree of spatial inhomogeneity of the barrier has to be taken into account. Numerous factors, intrisically linked to the interface atomic structure (semiconductor defects, contaminants, mixture of phases, grain boundaries, doping inhomogeneity and so on) [23,30,31], could be at the origin of spatial inhomogeneity of the Schottky barrier height.

To explain the effect of the inhomogeneity on the electrical characteristics of the metal/4H-SiC Schottky contacts, we invoked Tung's theory [23], which assumes small region areas ("patches") featuring low Schottky barrier height, embedded in a uniform background with higher Schottky barrier. In particular, this model was previously applied to explain the electrical behaviour in in several Schottky contacts to 4H-SiC [17,32,33].

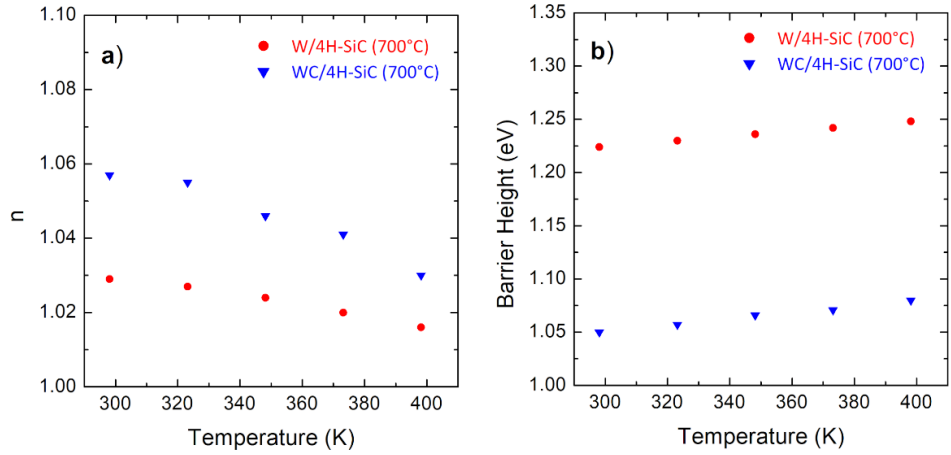

**Figure 7**. Temperature-dependence of *a)* ideality factor n and *b)* barrier height $\Phi_B$ for the W/4H-SiC and WC/4H-SiC Schottky contacts subjected to thermal annealing at 700 °C.

Accordingly, by considering the temperature-dependence of the saturation current as reported in eq. (2):

$$I_S = AA^*T^2 exp[(-q\phi_B/k_BT)] \qquad (2)$$

it is possible to extrapolate the "effective" values of barrier height (i.e. a barrier height averaged over all of the patches, lower than the $\Phi_B$ derived by TE analysis of room-temperature forward I-V curve) and the effective Richardson's constant [5]. These parameters were derived by a linear fit in the conventional Richardson's plot $\ln\frac{I_S}{T^2}$ *vs* $\frac{q}{k_BT}$, reported in Fig.8. In our cases, the linear fit provides an effective barrier height $\Phi_B^{eff} = 1.15$ eV and an effective Richardson' constant $A^*_{eff} = 9$ A×cm$^{-2}$×K$^{-2}$ for the W/4H-SiC contact, and $\Phi_B^{eff} = 0.96$ eV and $A^*_{eff} = 4.7$ A×cm$^{-2}$×K$^{-2}$ for the WC/4H-SiC one. It is common to find effective Richardson's constant two orders of magnitude lower that the expected value (146 A×cm$^{-2}$×K$^{-2}$ [27]), when local inhomogeneity occurs at the interface [27,32].

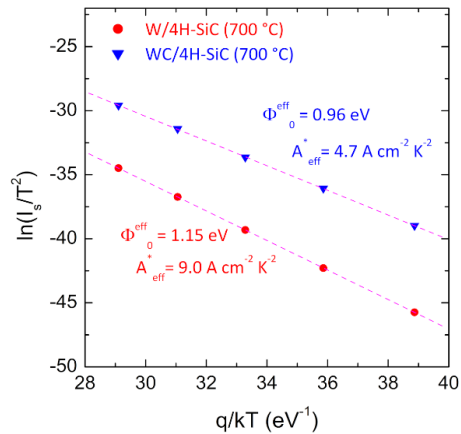

**Figure 8**. Conventional Richardson's plot $\ln(I_S/T^2)$ *vs* $1/k_BT$ for the W/4H-SiC and WC/4H-SiC Schottky contacts annealed at 700 °C. The effective barrier height $\Phi_B^{eff}$ and effective Richardson's constant $A^*_{eff}$ are reported in the graph.

In fact, the y-intercept in the Richardson's plot gives the product A×A*. Hence, a significant low value of this product is estimated if the "active area", namely the contact area effectively interested by the current transport, is much smaller than the entire area of the contact as it is assumed in the TE model.

A practical method to quantify the local Schottky barrier inhomogeneity, is to compare the $nk_BT$ curve with the ideal case (n=1) in a characteristic plot $nk_BT$ vs $k_BT$, as illustrated in Fig. 9a. The experimental data can be fitted by a straight line parallel to that of the ideal case, that is a behaviour commonly referred as "T$_0$ anomaly" [34]. Accordingly, the temperature dependence of the ideality factor can be expressed as:

$$n = 1 + T_0/T \qquad (3)$$

and from the linear fit of the nk$_B$T *vs* k$_B$T plot, we obtained T$_0$ = 15K for W/4H-SiC and 34K for WC/4H-SiC contact.
The ideal barrier height $\Phi_B^0$ of the uniform background surrounding the low-barrier patches, can be determined as the limit for n=1 in a barrier height $\Phi_B$ versus ideality factor n plot [27,35], as that presented in Fig. 9b. In particular, we extrapolated a barrier height $\Phi_B^0$=1.28 eV and 1.11 eV for the W/4H-SiC and WC/4H-SiC contact, respectively.

In general, according to Tung's model, the inhomogeneous contact is described by a distribution of "patches" characterized by a "patch parameter" γ, which is related to the surface potential Δ and radius R$_0$ (assuming circular area of the patch), through the relation [34]

$$\gamma = 3(\Delta R_0^2/4)^{1/3} \qquad (4)$$

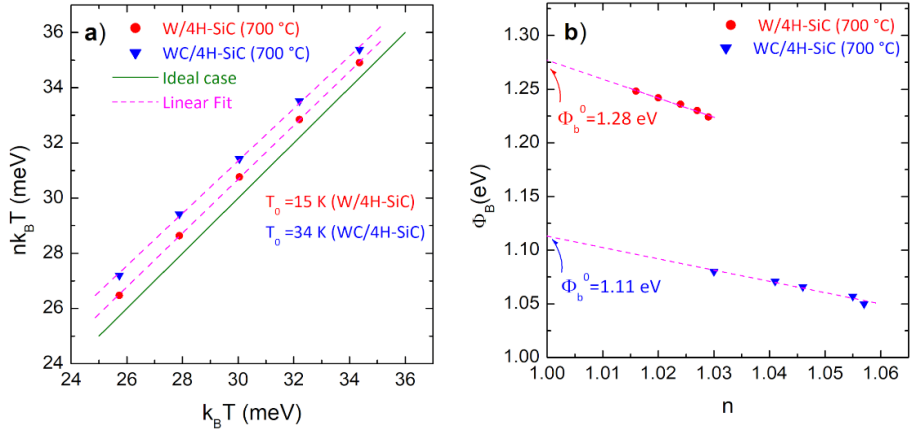

**Figure 9**. For W/4H-SiC and WC/4H-SiC contacts annealed at 700°C: *a)* plot of nk$_B$T *vs* k$_B$T for determing the T$_0$ anomaly from n=1+T$_0$/T expressing the deviation from the ideal case with n=1 (depicted with a solid green line); *b)* plot of Schottky barrier height $\Phi_B$ *vs* ideality factor n, with the homogenoeus barrier $\Phi_B^0$ extrapolated by a linear fit at n=1.

Assuming for simplicity that all the patches are characterized by the same γ parameter, the deviation of the effective barrier from the uniform barrier height is related to the patch parameter by

$$\Phi_B^0 - \Phi_B^{eff} = \gamma \left(\frac{V_{bb}}{\eta}\right)^{1/3} \qquad (5)$$

where V$_{bb}$ is the band bending potential and η=ε$_s$/qN$_D$. Hence, a γ parameter of 9.6×10$^{-5}$ and 1.1×10$^{-4}$ V$^{1/3}$cm$^{2/3}$ was derived for the W/4H-SiC and WC/4H-SiC contact, respectively.
From these values, an effective area A$_{eff}$ (that is the active area of a single low barrier patch effectively involved in the current transport) can be derived by

$$A_{eff} = \frac{4\pi}{9k_BT}\gamma \left(\frac{\eta}{V_{bb}}\right)^{2/3}. \qquad (6)$$

From this calculation, the single low barrier patch had an area of $2.85 \times 10^5$ nm$^2$ for the W/4H-SiC, lower than that obtained for the WC/4H-SiC of $3.27 \times 10^5$ nm$^2$, corresponding to a radius of 301 e 323 nm, in the assumption of a circular shape of the patches. This estimation indicates that the Schottky barrier inhomogeneity occurs at a nanometre scale.

In the real case, an inhomogeneous barrier is characterized by a Gaussian distribution of patch parameters $\gamma_i$. The standard deviation of the $\gamma$ parameter $\sigma_\gamma$ is associated with the $T_0$ parameter by the relation [34,36]:

$$T_0 = q\,\sigma_\gamma^2 / 3k_B\,\eta^{2/3} V_{bb}^{1/3} \quad (7)$$

Hence, considering the values of $T_0$=15 and 34 K obtained in our case (Fig. 8a), we calculated the values of $\sigma_\gamma$, $5.92 \times 10^{-5}$ and $8.87 \times 10^{-5}$ V$^{1/3}$cm$^{2/3}$ for the W/4H-SiC and WC/4H-SiC contact, respectively. The finding of a slightly narrower distribution for the W/4H-SiC is consistent with the observation of a lower ideality factor for this case.

In order to have a more complete picture of the electrical behaviour of the Schottky diodes under investigation, we performed an I-V-T characterization under reverse bias. The analysis of the J-V characteristics under reverse bias for the W/4H-SiC and WC/4H-SiC contacts annealed at 700 °C is reported in Figs. 10a and 10b, respectively. Typically, for 4H-SiC Schottky diodes a level of reverse current higher than that predicted by the thermionic emission theory is observed, justifiable by taking into account a tunnelling contribution to the reverse leakage current [37,38]. According to the thermionic field emission (TFE) theory, the relationship between reverse current density $J$ and voltage $V_R$ is [39]:

$$J = A^* T^2 \sqrt{\frac{q\pi E_{00}}{kT}} \sqrt{V_R + \frac{\phi_B}{\cosh\left(\frac{qE_{00}}{kT}\right)^2}}\, exp\left(-\frac{\phi_B}{E_1}\right) exp\left(\frac{V_R}{E_2}\right) \quad (12)$$

where $E_{00} = (h/4\pi)\sqrt{N_D/m^*\varepsilon_{SiC}}$, $E_1 = E_{00} \times \tanh(qE_{00}/kT)^{-1}$ and $E_2 = E_{00} \times \left(qE_{00}/kT - tanh(qE_{00}/kT)\right)^{-1}$ and with $h$ the Planck constant, $m^*$ the effective mass of electron and $\varepsilon_{SiC}$ the dielectric constant of the semiconductor.

Figs. 10a and Fig 10b reports the experimental and simulated reverse leakage current curves for different measurement temperatures. In particular, we considered only the measurement temperatures where the leakage current was well above the sensitivity limit of our experimental set-up.

As can be seen, the leakage current increased as function of the reverse bias and temperature increasing. Moreover, for the W/4H-SiC Schottky contact, the leakage current is systematically lower that that observed for the WC/4H-SiC contact, as could be expected considering the higher value of Schottky barrier in W/4H-SiC contact.

Noteworthy, for each measurement temperature, the barrier height derived from the temperature-dependent analysis of the forward characteristics has been used to simulate the TFE curves. A very good agreement between the experimental curves and the curves simulated with the TFE model was observed.

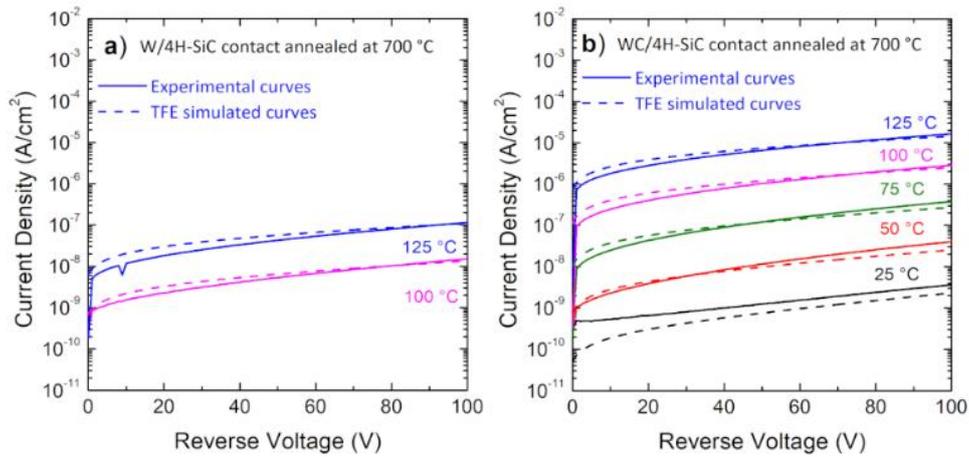

**Figure 10**. Experimental reverse current density-voltage (J-V) curves for *a)* W/4H-SiC and *b)* WC/4H-SiC Schottky contacts acquired at variuos measuerement tempertaures (solid-lines). For both cases, the curves were compared to simulated characteristics according to the TFE theory (dashed-lines).

## 4. Conclusions

In this work, we have studied the electrical properties of the tungsten (W) and tungsten carbide (WC) Schottky contacts on 4H-SiC in the as-deposited case and following their evolution in samples subjected to different annealing temperature (increasing from 475°C to 700 °C).

Both contacts showed a nearly ideal behaviour with low value of ideality factor up to annealing at 700°C. Moreover, while the W/4H-SiC contact presented an increase of the barrier with the increase of the annealing temperature, the WC/4H-SiC diode is characterized by a slight reduction of the barrier up to annealing temperature of 700 °C. The narrow statistical distribution of the barrier height values measured on a set of equivalent diodes over the wafer confirmed the good reproducibility of these Schottky barriers.

The temperature dependent I–V characterization of the diodes annealted at 700°C revealed an increase of the barrier height and a decrease of the ideality factor as function of the measurement temperature increase, which could be explained by the Tung's model assuming a lateral inhomogeneity of the contacts. In this regard, the forward characteristics can be described by the thermionic-emission (TE) theory with an effective barrier $\Phi_B^{eff}$ of 1.15 and 0.96 eV and a homogeneous barrier $\Phi_B^0$ of 1.28 and 1.11 eV for the W/4H-SiC and WC/4H-SiC contact, respectively.

Under reverse bias, the I-V characteristics can be described by the thermionic field emission (TFE) formalism by including the temperature dependence of the barrier height, as derived in forward bias analysis. This characterization supplies a view on the electrical behaviour of W/4H-SiC and WC/4H-SiC contacts subjected to various annealing temperature, and can be useful to evaluate their employment in as operative devices.

## Acknowledgements

The authors would like to acknowledge G. Greco, F Giannazzo, and M. Saggio (STMicroelectronics) for fruitful discussions. S. Di Franco (CNR-IMM) is acknowledged for his technical support during device fabrication. This work was partially supported by the European ECSEL JU project REACTION (Grant Agreement No. 783158), using the facilities of the Italian Infrastructure Beyond-Nano. The authors thank also the Bionanotech Research and Innovation Tower (BRIT) laboratory of University of Catania (Grant no. PONa3_00136 financed by the Italian Ministry for Education, University and Research, MIUR) for the diffractometer facility.